\documentclass[letterpaper, twocolumn, prl]{revtex4}
\usepackage{graphicx, amsmath, amssymb, bm, titlesec, epstopdf}

\begin{document}
\title{Angular Lens}

\author{Rishabh Sahu$^{1, \dagger}$, Swati Chaudhary$^{1, \dagger, \#}$, Kedar Khare$^2$, Mishkatul Bhattacharya$^3$, Harshawardhan Wanare$^1$, and Anand Kumar Jha$^1$}

\email{akjha9@gmail.com }

\thanks{$\#$ Presently at California Institute of Technology, Pasadena, CA, USA}

\thanks{$\dagger$ Both the authors contibuted equally to the work}

\affiliation{ $^1$  Department of Physics, Indian Institute of Technology Kanpur, Kanpur, UP 208016, India \\
$^2$Department of Physics, Indian Institute of Technology Delhi, Hauz Khas, New Delhi 110016, India \\
$^3$ School of Physics and Astronomy, Rochester Institute of Technology, Rochester, New York 14623, USA}

\date{\today}

\begin{abstract}
We propose a single phase-only optical element that transforms different orbital angular momentum (OAM) modes into localized spots at separated angular positions on a transverse plane. We refer to this element as angular lens since it separates out OAM modes in a manner analogous to how a converging lens separates out transverse wave-vector modes at the focal plane. We also simulate the proposed angular lens using a spatial light modulator and experimentally demonstrate its working. Our work can have important implications for OAM-based classical and quantum communication applications.
\end{abstract}

%\doi{\url{http://dx.doi.org/10.1364/optica.XX.XXXXXX}} % REPLACE WITH CORRECT OCIS CODES FOR YOUR ARTICLE, MINIMUM OF TWO; Avoid using the OCIS codes for “General” or “General science” whenever possible.
%For a complete list of OCIS codes, visit: https://www.osapublishing.org/oe/submit/ocis/

\maketitle

\section{Introduction}

It is known that the transverse position and the transverse wave-vector bases form a two-dimensional Fourier transform pair and that a converging lens is a phase-only optical element that performs this Fourier transformation \cite{goodman, born&wolf}. Owing to this transformation property of a lens, optical modes characterized by different transverse wave-vectors get mapped onto separated localized spots on a transverse plane after passing through a lens. When the aperture-size of the lens is infinite, the localized spots take the form of two-dimensional Dirac-delta functions and the  wave-vector separation is said to be perfect. However, with a finite aperture-size lens, this separation is imperfect and its degree characterizes the resolving power of the lens.

It is now also known that optical modes having an $e^{-i\ell\phi}$ phase profile can carry $\ell\hbar$ orbital angular momentum (OAM) per photon \cite{allen1992pra}. Here $\phi$ is the angular position and $\ell$ is referred to as the aziumthal mode index or the OAM mode index. This feature of OAM modes has made them extremely important for communication and computation protocols, in terms of system capacity \cite{wang2012natphot,bozinovic2013science,yan2014natcom}, security
\cite{karimipour2002pra,cerf2002prl,nikolopoulos2006pra},
transmission bandwidth \cite{fujiwara2003prl,cortese2004pra}, gate
implementations \cite{ralph2007pra,lanyon2009natphy},
supersensitive measurements \cite{jha2011pra2} and fundamental
tests of quantum mechanics
\cite{kaszlikowski2000prl,collins2002prl,vetesi2010prl,
leach2009optexp}. However, one major challenge in implementing OAM-based protocols is the efficient separation and detection of OAM-modes. The earliest efforts at separating OAM modes were based on using a phase-only hologram, either thin \cite{mair2001nature, gibson2004optexp}  or thick \cite{gruneisen2011njp}. But these methods  turned out to be quite inefficient and are not suitable at single photon levels. Later, techniques based on concatenated Mach-Zehnder interferometers \cite{leach2002prl, leach2004prl} and rotational Doppler shift were proposed \cite{vasnetsov2003optlett, courtial1998prl}. Although these techniques are in principle 100$\%$ efficient even at the single-photon level, it is extremely difficult to implement them for more than a few modes. More recently, there have been efforts \cite{berkhout2010prl, lavery2012optexp, lightman2017optica} based on log-polar mapping \cite{bryngdahl1974josa, bryngdahl1974optcomm} that can work with more modes and also at the single-photon level. However, these recent methods involve several elements and are quite cumbersome for optical fields containing several OAM modes. Therefore, the existing methods for separating out OAM modes are either inefficient or unsuitable at single-photon levels, or involve multiple elements for their implementation.

In this article, we propose and demonstrate a single phase-only optical element that separates out OAM modes into localized spots in much the same way as a converging lens separates out transverse wave-vector modes. We refer to this element as an ``angular lens'' and show that it provides a natural way of separating out OAM modes and can not only work with a large number of incoming modes but also at the single-photon level.

\section{Angular Lens: The phase transformation function and its  action on OAM modes}
\begin{figure}
\includegraphics{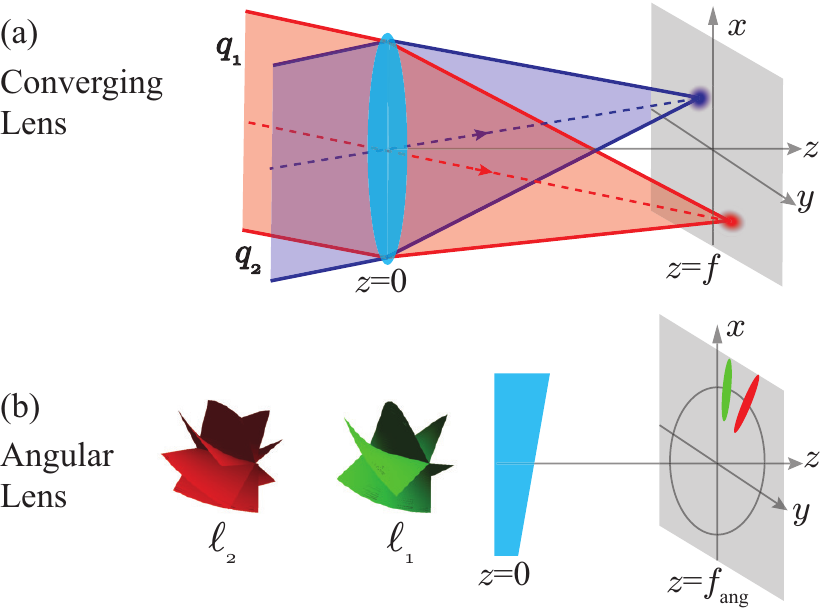}
\caption{(a) The schematic illustration of the working of a converging lens. The transverse wave-vector modes ${\bm 	q_1}$ and ${\bm q_2}$ with phase profiles $e^{i{\bm q_1}\cdot\bm\rho}$ and  $e^{i{\bm q_2}\cdot\bm\rho}$  get localized at separate spatial locations. (b) The schematic illustration of the working of our proposed angular lens. The OAM modes $\ell_1$ and $\ell_2$ with phase profiles $e^{-i\ell_1\phi}$ and $e^{-i\ell_2\phi}$ get localized at separate angular positions.}
\label{fig1}
\end{figure}
Figure \ref{fig1}(a) illustrates how a converging lens separates out different transverse wave-vector modes. The phase transformation function $T(x, y)$ of a thin converging lens within the paraxial approximation is given by (see Section 5.2 of Ref.~\cite{goodman}):
$T(x, y)=\exp[-\frac{ik}{2f}(x^2+y^2)]$,
where $f$ is the focal length of the lens and where $k=2\pi/\lambda$ with $\lambda$ being the wavelength of light. The lens transforms the transverse wave-vector modes ${\bm 	q_1}$ and ${\bm q_2}$ with phase profiles $e^{i{\bm q_1}\cdot\bm\rho}$ and  $e^{i{\bm q_2}\cdot\bm\rho}$, where $|{\bm q_1}|, |{\bm q_2}| \ll k$, into localized spots on a transverse plane kept at $z=f$. Figure \ref{fig1}(b) is the schematic illustration of the expected working of an angular lens. We now want to find out the phase transformation function of a lens that performs as depicted in Fig.~\ref{fig1}(b). 

\begin{figure}[t!]
\centering\includegraphics[width=\linewidth]{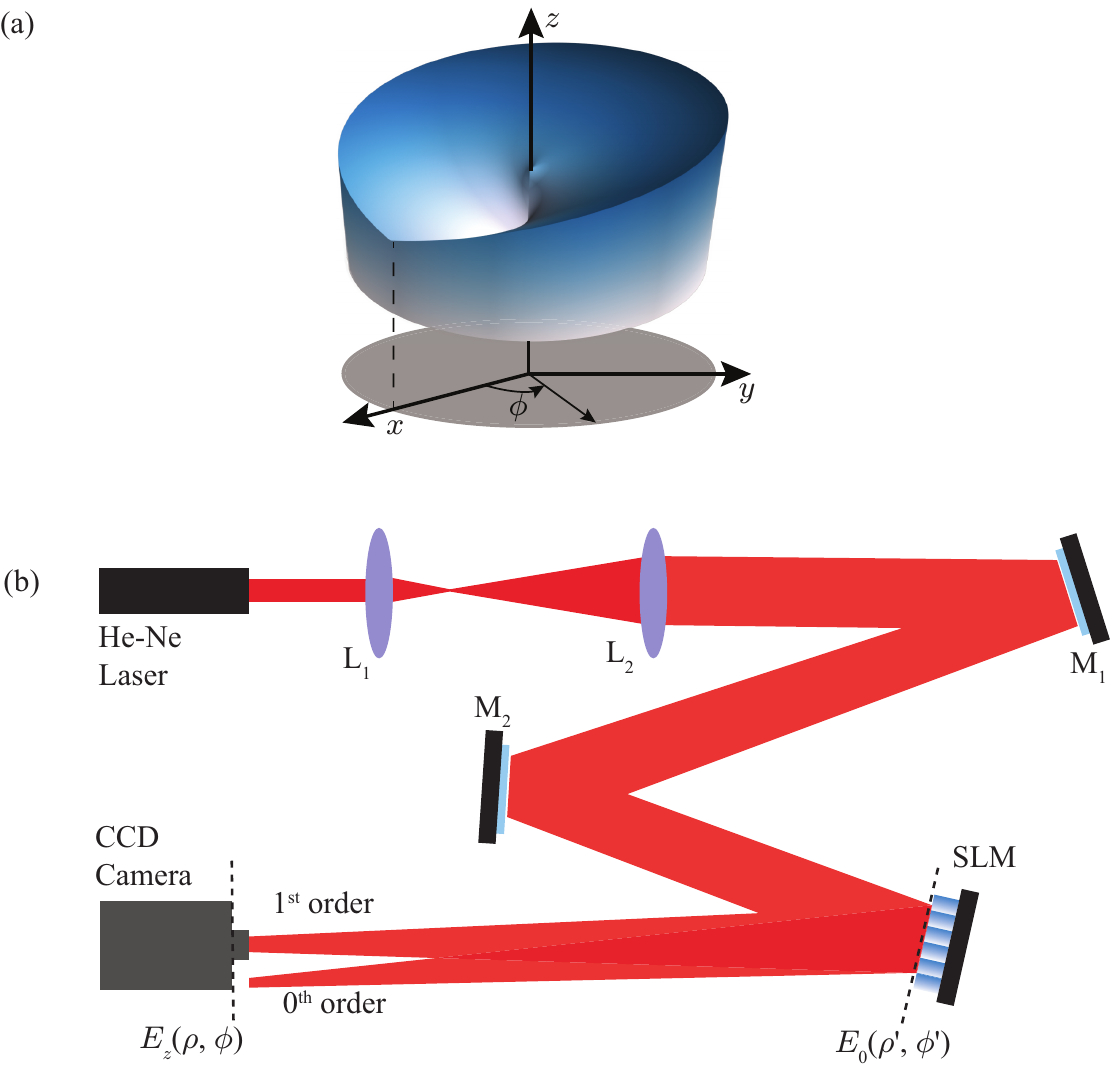}
\caption{(a) The thickness function of the proposed angular lens. (b) The experimental Setup. SLM is spatial light modulator, L stands for a converging lens, and M stands for a mirror.}
\label{fig3}
\end{figure}

We begin by noting that the angular-position and the orbital angular momentum (OAM) bases form a Fourier transform pair in much the same way as the transverse position and transverse wave-vector bases do \cite{yao2006opex, jha2008pra2, jha2010prl}. Therefore, it is natural to expect the transformation function of an angular lens to have a quadratic dependence on $\phi$ just as the transformation function of a converging lens has quadratic dependences on $x$ and $y$. However, unlike the transverse position coordinates ($x$, $y$) the cylindrical coordinates ($\rho, \phi$) do not form a two-dimensional Fourier pair. Therefore, it is not straightforward to arrive at an analogous functional dependence on $\rho$. Nevertheless, we take a hint from Ref. \cite{arlt2001jmo}, in which it was shown that an axicon, which has a transformation function given by $e^{i\beta\rho}$ with $\beta$ being a constant, transforms a Laguerre-Gaussian mode into an ultranarrow annulus. With this hint, we take the following as the transformation function $T_{\rm ang}(\rho, \phi)$ of our proposed angular lens:  
\begin{align}
T_{\rm ang}(\rho, \phi)=\exp\left[-i(\alpha\phi^2-\beta\rho)\right].
\label{eqn:ang_lens_phase}
\end{align} 
Here $\alpha,\beta$ are two constants and $\phi\in[-\pi,\pi]$ and $\rho\in[0,\infty]$. The thickness function corresponding to the phase transformation function has been plotted in Fig.~\ref{fig3}(a).

\begin{figure}
\includegraphics[width=\linewidth]{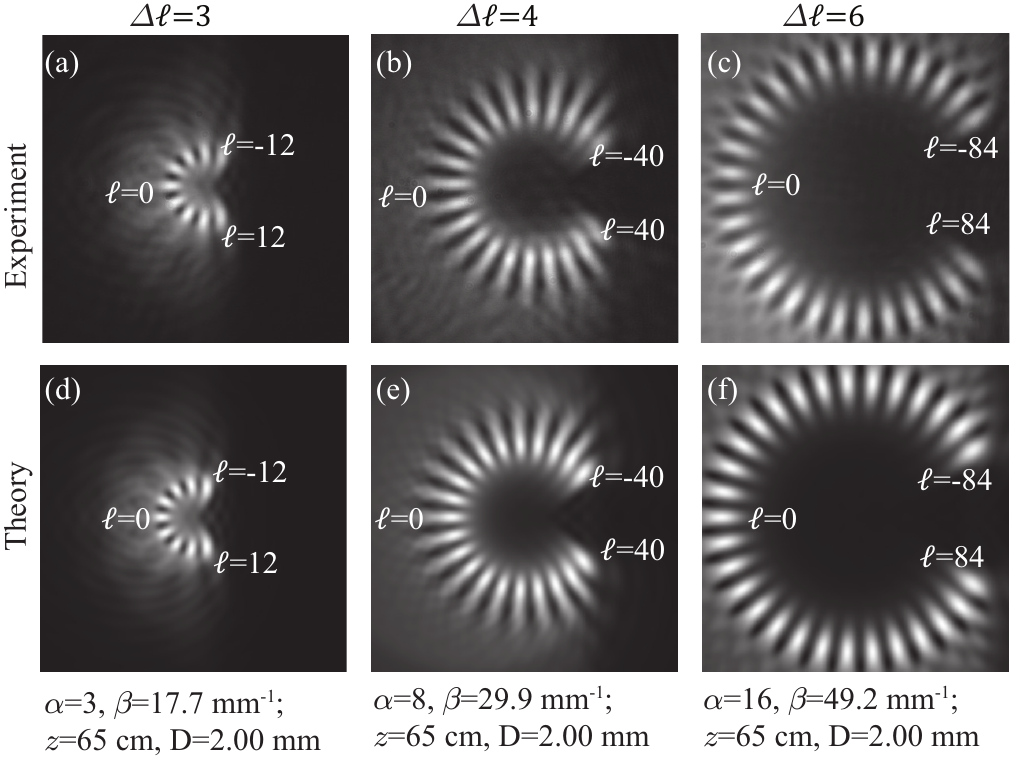}
\caption{Transformation of constant-intensity OAM modes by the angular lens. (a), (b), and (c) are the combined intensity patterns for various $\ell$ observed at $z=65$ cm. (d)-(f) are the theoretical diffraction patterns as obtained by numerically evaluating the integral in Eq.~(\ref{fresnel_polar}) for the same set of parameters. The screen size in all the above plots is 6.27 mm $\times$ 6.27 mm. A constant background of 250 counts has been subtracted from the each CCD camera pixel.}
\label{fig4}
\end{figure}

We now illustrate the workings of our proposed angular lens using the experimental setup shown in Fig.~\ref{fig3}(b). An angular lens with the transformation function $T_{\rm ang}(\rho, \phi)$ given by Eq.~(\ref{eqn:ang_lens_phase}) is placed at $z=0$ and an input field with amplitude $E_{z=0}(\rho', \phi')$ at $z=0$ is incident on it. The field amplitude $E_{z=z}(\rho, \phi)$ at $z$ is given by the Fresnel diffraction integral \cite{goodman}:
\begin{align}
E_{z=z}(\rho,\phi)&=\frac{e^{ikz}}{i\lambda
z}e^{i\tfrac{k}{2z}\rho^2}\int_0^\infty\int_0^{2\pi} E_{z=0}(\rho',\phi') \notag \\ &\times T_{\rm ang}(\rho, \phi) 
e^{i\tfrac{k}{2z}\rho'^2}   e^{-i\tfrac{k\rho\rho'}{z}\cos(\phi-\phi')}\rho' d\rho'
d\phi'.\label{fresnel_polar}
\end{align}
First of all we investigate the transformation properties of our angular lens for input field modes given by $E_{z=0}(\rho', \phi')=e^{-i\ell\phi'}$. Such modes have constant transverse intensity at $z=0$. In our experiment, we generate these modes one by one in a sequential manner, and for each generated mode with a given OAM mode index, we measure the diffracted intensity pattern at $z$. The modes are generated by first expanding our continuous-wave He-Ne laser beam to be 1-cm wide. We then diffract this laser beam from the spatial light modulator (SLM) kept at $z=0$ after putting an appropriate phase pattern on it \cite{mair2001nature}. We also put a circular aperture of diameter $D=2$ mm onto the SLM so that only a small circular portion of the incoming laser beam undergoes diffraction and thus, to a good approximation, the intensity in the circular portion can be taken as constant. The transformation function corresponding to the angular lens at $z=0$ is also simulated using the same SLM. With an SLM, the diffraction pattern corresponding to the transmission function simulated on it is observed at the first diffraction order; the zeroth diffraction order of an SLM contains mostly the reflected portion of the incoming field and does not contain much information about the transmission function simulated on the SLM \cite{mair2001nature,arrizon2007josaa}. Therefore, we record the transverse intensity at the first SLM diffraction order using a CCD camera placed at $z$. The parameters $\alpha$  and $\beta$ are electronically changed in order to simulate different lenses. By changing $\ell$ in a sequential manner, we generate a range of OAM modes and for each of these modes we measure the diffracted intensity pattern at $z$. After collecting the intensity patterns for various different values of $\ell$, we plot the combined intensity patterns.

Figure \ref{fig4} shows the combined intensity patterns observed by the CCD camera kept at $z=65$ cm for a range of OAM modes with various $\ell$ and with separation $\Delta\ell$. Figures \ref{fig4}(a), \ref{fig4}(b), and \ref{fig4}(c) show the combined intensity patterns corresponding to $\alpha$ equal to 3, 8, and 16, respectively. In order to consider the two modes as separated we adapt the following resolution criterion: if in the  combined two-dimensional intensity plot at $z$, the ratio of the intensity at the minimum located in between the two maxima and that at the maxima is less than about 0.3 then the two modes are resolved. This criterion is much stringent than that of Rayleigh, which allows for ratios up to 0.811. The value of $\beta$ was obtained by optimizing it for a given value of $\alpha$ and $z$ such that we obtain a localized diffraction pattern and the resolution criterion is satisfied. For the three $\alpha$ values, the optimized values of $\beta$ were found to be  $17.7$ mm$^{-1}$, $29.9$ mm$^{-1}$, and $49.2$ mm$^{-1}$, respectively. We find that different OAM modes get transformed into diffraction patterns localized at separate angular positions and that as $\alpha$ increases the mode separation $\Delta\ell$ that could be resolved increases as well. For the three $\alpha$ values, we find that modes with separation $\Delta\ell$ equal to 3, 4, and 6 could be resolved. Further, we find that as $\alpha$ increases, the range of modes that can be transformed into localized functions also increases. Figures \ref{fig4}(d)-\ref{fig4}(f) show the corresponding theoretical diffraction patterns as obtained by numerically evaluating the integral in Eq.~(\ref{fresnel_polar}) for the same set of parameters. We find an excellent agreement between the theory and experiments.

We note that for the aperture size of $D=2$ mm, $\Delta\ell=3$ is the lowest value. For $\alpha=3$ and $D=2$ mm, $\beta$ cannot be optimized to satisfy the resolution criterion with $\Delta\ell=2$ or $\Delta\ell=1$. This, in fact, is a generic feature of optical elements having finite sizes. For example, a converging lens achieves perfect resolution only when the aperture size is infinite. With finite aperture-sizes, the resolving power of a lens remains limited \cite{goodman}. Similar limitation on resolution is also observed in the log-polar mapping based method for sorting OAM modes \cite{bryngdahl1974josa, bryngdahl1974optcomm}. We also note in Fig.~\ref{fig4} that there exists a limit on the maximum value of $\ell$ up to which the angular lens produces localized patterns. Beyond this maximum value the transformation function no longer produces localized patterns. In the figure, we have plotted the results only up to this maximum $\ell$ value. The maximum value of $\ell$ seems to scale as $\alpha$ in the sense that as $\alpha$ increases the maximum value of $\ell$ also increases in a linear manner.

In a converging lens the resolving power is decided by the size of the lens. In our proposed angular lens, the parameter $\alpha$ is playing an analogous role. Both the range of the mode and the minimum separation $\Delta\ell$ that could be resolved depends on the parameter $\alpha$. As $\alpha$ increases, the range of modes that can be resolved increases but the resolution decreases. The parameter $\beta$ is obtained by optimizing it for a given value of $\alpha$ and $z$ such that we obtain a localized diffraction pattern and the resolution criterion is satisfied. The parameter $\beta$ plays a somewhat analogous role as the focal length of a converging lens. 

\begin{figure}[t!]
\includegraphics[width=\linewidth]{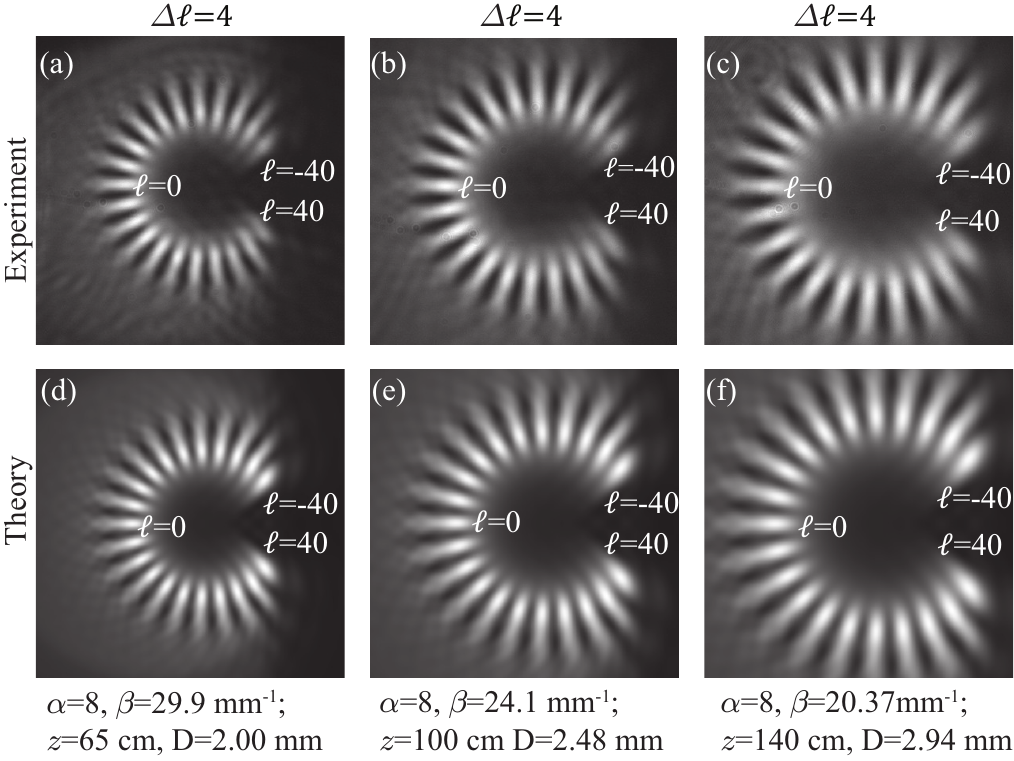}
\caption{Scaling of diffraction patterns with $z$. (a), (b), and (c) are the combined intensity patterns observed at three $z$ values. (d)-(f) are the theoretical diffraction patterns as obtained by numerically evaluating the integral in Eq.~(\ref{fresnel_polar}) for the same set of parameters. The screen size in all the above plots is 6.27 mm  $\times$ 6.27 mm. A constant background of 250 counts has been subtracted from the each CCD camera pixel.}
\label{fig5}
\end{figure}

In order to illustrate this analogous property of the $\beta$ parameter, we first recall how the focal length of a converging lens transforms a given input field. We know that for a given input field the focal intensity patterns due to lenses with different focal lengths remain the same except for an overall scaling of the pattern. In the proposed angular lens, the $\beta$ parameter shows an analogous scaling property for fixed $\alpha$ values. In order to show this scaling, let us consider two angular lenses with the same $\alpha$ but with $\beta$ parameters being equal to $\beta_1$ and $\beta_2$ and the aperture size $D$ being equal to $D_1$ and $D_2$, respectively. Let us assume that with $\beta_1$ and $D_1$ the angular lens produces the optimized diffraction pattern at $z=z_1$. The field amplitude $E_{z=z_1}^{(1)}(\rho,\phi)$ in this case can be written using Eq.~(\ref{fresnel_polar}) as 
\begin{align}
E_{z=z_1}^{(1)}(\rho,\phi)&=\frac{e^{ikz_1}}{i\lambda
z_1}e^{\tfrac{ik\rho^2}{2z_1}}\int_0^{D_1/2}\int_0^{2\pi} E_{z=0}(\phi') \notag \\ &\times T_{\rm ang}(\rho, \phi)
e^{i\tfrac{k}{2z_1}\rho'^2}   e^{-i\tfrac{k\rho\rho'}{z_1}\cos(\phi-\phi')}\rho' d\rho'
d\phi'.\label{fresnel_polar2}
\end{align}
We at once see that since the input field amplitude depends only on $\phi'$ the functional form of the intensity due to the first angular lens at $z_1$ is equal to that due to the second lens at $z_2$, that is, 
\begin{align}
|E_{z=z_2}^{(2)}(\rho,\phi)|^2&=|E_{z=z_1}^{(1)}(\rho/a,\phi)|^2,\notag \\ &\quad {\rm if} \quad
z_2=a^2 z_1, \ \beta_2=\frac{\beta_1}{a} \ {\rm and} \ D_2=\frac{D_1}{a}, \label{scaling}
\end{align}
for a given constant value of $a$. We thus find that if both $\beta$ and $D$ are decreased by a factor of $a$, one obtains a radially scaled up version of the same diffraction pattern at a propagation distance $z$ that is $a^2$ times larger. Figure \ref{fig5} shows the experimental and theoretical results illustrating this analogous focusing property of the angular lens. The diffraction pattern in Fig.~\ref{fig5}(a) is for a lens with $\alpha=8$ and $D=2$ mm, and the $\beta$ parameter was obtained by optimizing it such that we obtain a localized diffraction pattern and the resolution criterion is satisfied. For the results in Figs.~\ref{fig5}(b) and \ref{fig5}(c), we chose $z$ to be $100$ cm and $140$ cm respectively and the corresponding $\beta$ and the size of the lens were chosen simply using the scaling in Eq.~(\ref{scaling}), without any optimization. Figures \ref{fig5}(d)-\ref{fig5}(f) are the theoretical diffraction patterns as obtained by numerically evaluating the integral in Eq.~(\ref{fresnel_polar}) for the same set of parameters. 
\begin{figure}
\includegraphics[scale=1]{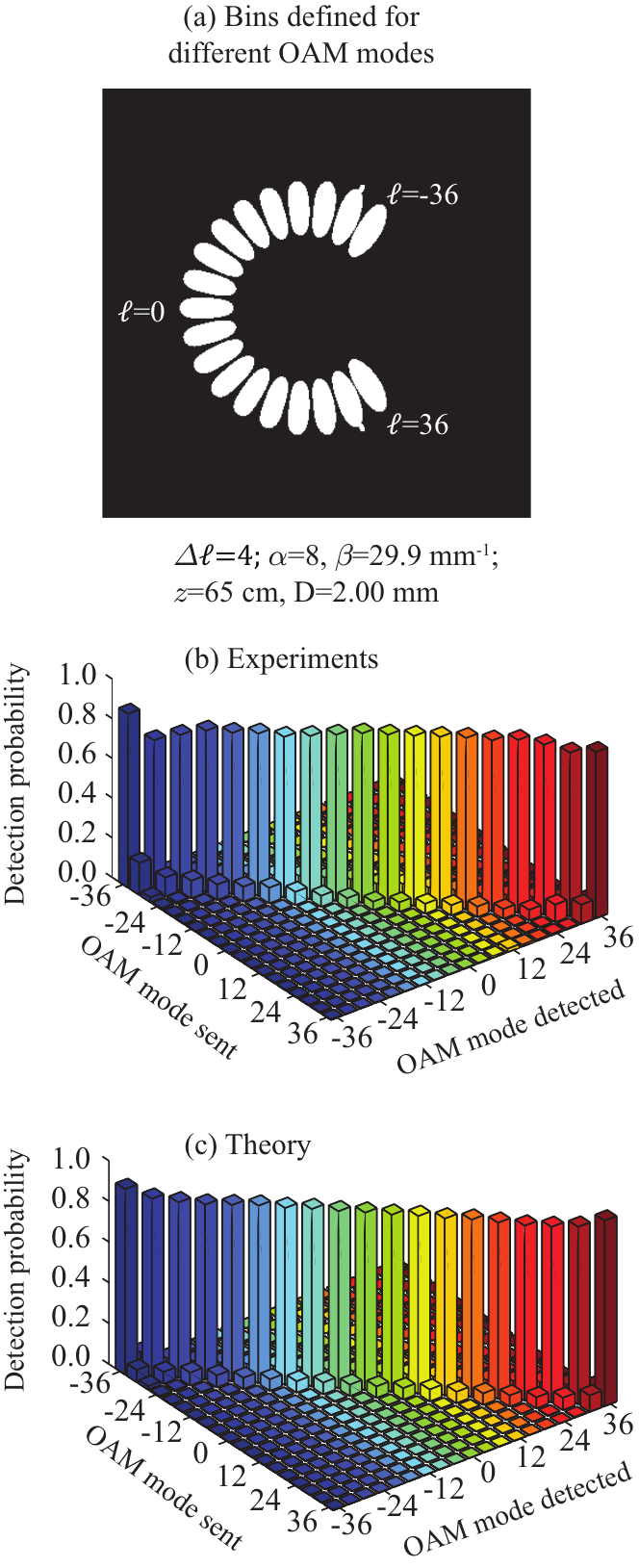}
\caption{ Cross-talk analysis. (a) The theoretically defined bins for different OAM modes. (b) Experimental detection probability of the 19 input OAM modes. (c) The theoretical detection probability of the 19 input OAM modes.}
\label{fig7}
\end{figure}

Our results so far have illustrated how our proposed angular lens transforms different OAM modes into localized spots at separated angular positions on a transverse plane. We have shown that our angular lens separates out OAM modes in a manner analogous to how a converging lens separates out transverse wave-vector modes at the focal plane. We next quantify the resolving power of our angular lens in terms of its use as an OAM sorter. For this purpose, we adopt the overall procedure of Refs.~\cite{mirhosseini2013natcomm, bozinovic2013science, huang2015scirep} and calculate the cross-talk for the set of lens parameters reported in figs.~\ref{fig4}(b) and \ref{fig4}(c). First, we divide the detection area on the CCD into 19 non-overlapping spatial bins so as to define a detection bin for each OAM mode with index ranging from $\ell=-36$ to $\ell=36$ with separation $\Delta\ell=4$. In order to define these bins, we have used the diffracted intensity expression of Eq.~(\ref{fresnel_polar}) for the given set of lens parameters and labeled a given set of pixels as one bin if the set of pixels have at least 25\% of the intensity of the most-intense pixel. The pixels having less than 25\% intensity do not define any bin. The spatial bins defined this way have been plotted in fig.~\ref{fig7}(a). We then send a known OAM mode through our system and record the intensity in each of the 19 spatial bins. We repeat this for all the 19 input OAM modes and plot the detection probability for each spatial bin in fig.~\ref{fig7}(b). Figure \ref{fig7}(c) shows the theoretical detection probability for each spatial bins. The cross-talk for a given mode has been defined as the fraction of the input intensity collected in spatial bins other than the one meant for the given mode. The experimental cross-talk averaged over all the 19 modes turns out to be 16.5\%. The theoretically calculated average cross-talk come out to be 12.5\%.  We note that, in the context of log-polar mapping based method \cite{berkhout2010prl, lavery2012optexp, osullivan2012optexp, mirhosseini2013natcomm}, when the method is used in combination with the idea of beam copying \cite{romero2007josaa} modes with $\Delta\ell=1$ \cite{osullivan2012optexp, mirhosseini2013natcomm} can be separated with less than 10\% cross-talk. We believe that similar beam-copying techniques can also be employed to enhance the resolving power of our angular lens. Moreover, as opposed to the log-polar based methods, which require several elements for its implementation and are thus limited by the severe transmission loss \cite{mirhosseini2013natcomm} in the system, our angular lens is a single phase-only element and so when realized using a single glass element, instead on of an SLM, the transmission loss can be made negligibly small.

\begin{figure}
\includegraphics[width=\linewidth]{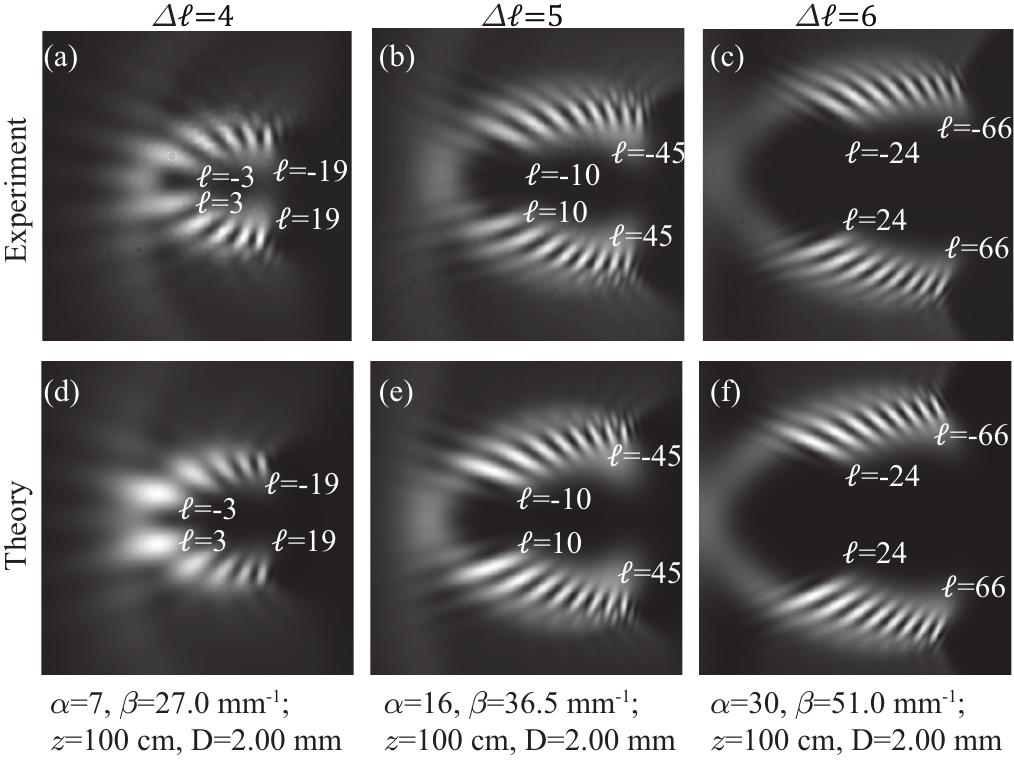}
\caption{Transformation of LG modes by the proposed angular lens. (a), (b), and (c) are the combined intensity patterns for various $\ell$ with $p=0$ observed at $z=100$ cm. (d)-(f) are the theoretical diffraction patterns as obtained by numerically evaluating the integral in Eq.~(\ref{fresnel_polar}) for the same set of parameters. The screen size in all the above plots is 5.63 mm $\times$ 5.63 mm. A constant background of 170 counts has been subtracted from the each CCD camera pixel.}
\label{fig6}
\end{figure}

\section{Action of angular lens on LG modes}

Next, we study the action of our angular lens on the Laguerre-Gaussian (LG) modes, which are the exact propagating solutions of the paraxial Helmholtz equation and are denoted by two indices, $\ell$ and $p$. The index $\ell$ decides the OAM content and the index $p$ decides the radial intensity distribution. We produce LG modes using the method by Arriz\'on \emph{et al.} \cite{arrizon2007josaa}. Figure \ref{fig6} shows the combined intensity patterns observed by the CCD camera kept at $z=100$ cm when a range of LG modes with $p=0$ and with separation $\Delta\ell$ were sequentially incident on angular lenses having various sets of values for $\alpha$ and $\beta$. Figures \ref{fig6}(a), \ref{fig6}(b), and \ref{fig6}(c) show the combined intensity patterns corresponding to $\alpha$ being equal to 7, 16, and 30, respectively. As before, for a given value of $\alpha$, the value of $\beta$ was obtained by optimizing it such that we obtain a localized diffraction pattern and the resolution criterion is satisfied. The values of $\beta$ for the three $\alpha$ values were found to be  $27.0$ mm$^{-1}$, $36.5$ mm$^{-1}$, and $51.0$ mm$^{-1}$, respectively. The mode separation $\Delta\ell$ that could be resolved for the three $\alpha$ values, were 4, 5, and 6, respectively. Figures \ref{fig6}(d)-\ref{fig6}(f) show the theoretical diffraction patterns as obtained by numerically evaluating the integral in Eq.~(\ref{fresnel_polar}) for the same set of parameters. Although the proposed angular lens is able to separate out the LG modes, it does not have all the analogous feature as in the case of flat-intensity OAM modes. More specifically, we do not see the same scaling as is seen in the case of constant-intensity OAM modes through Eq.~(\ref{fresnel_polar2}) and Eq.~(\ref{scaling}). This is because in this case the input field amplitude depends on both $\rho'$ and $\phi'$ and therefore Eq.~(\ref{fresnel_polar2}) does not show the same scaling.

\section{Summary}

In conclusion, we have proposed a single phase-only optical element that can transform different OAM modes into localized patterns at separated angular positions on a transverse plane. Using an SLM, we have experimentally demonstrated the working of our proposed angular lens for two different types of OAM modes. For constant-intensity OAM modes, our angular lens works in a manner analogous to how a converging lens works for transverse wave-vector modes. Even for the LG modes, our proposed angular lens is able to separate out the modes based on their OAM mode index. In several situations there are techniques that are employed to increase the resolving power of an optical element beyond its usual diffraction limit. For example, in the context of log-polar mapping based method \cite{berkhout2010prl, lavery2012optexp, osullivan2012optexp, mirhosseini2013natcomm}, it was shown that when the method is used in combination with the idea of beam copying \cite{romero2007josaa} it can separate out modes with $\Delta\ell=1$ \cite{osullivan2012optexp, mirhosseini2013natcomm}. We believe that similar techniques can also be employed to enhance the resolving power of our angular lens. Since the proposed angular lens is purely a phase-only element and works at any light level, we expect our work to have several important implications for OAM-based communication protocols in both classical \cite{wang2012natphot,bozinovic2013science,yan2014natcom} and quantum domains \cite{karimipour2002pra,cerf2002prl,nikolopoulos2006pra,
fujiwara2003prl,cortese2004pra,ralph2007pra,lanyon2009natphy,
jha2011pra2, kaszlikowski2000prl,collins2002prl,vetesi2010prl,
leach2009optexp}.

\section*{Funding}
The initiation
grant no. IITK /PHY /20130008 from Indian Institute of Technology
(IIT) Kanpur, India; The research grant no.
EMR/2015/001931 from the Science and Engineering Research Board
(SERB), Department of Science and Technology, Government of India; The grant no. 1454931 from the Directorate for Mathematical and Physical Sciences, National Science Foundation, USA.

\bibliography{Master_Ref1}

\begin{thebibliography}{39}
\expandafter\ifx\csname natexlab\endcsname\relax\def\natexlab#1{#1}\fi
\expandafter\ifx\csname bibnamefont\endcsname\relax
  \def\bibnamefont#1{#1}\fi
\expandafter\ifx\csname bibfnamefont\endcsname\relax
  \def\bibfnamefont#1{#1}\fi
\expandafter\ifx\csname citenamefont\endcsname\relax
  \def\citenamefont#1{#1}\fi
\expandafter\ifx\csname url\endcsname\relax
  \def\url#1{\texttt{#1}}\fi
\expandafter\ifx\csname urlprefix\endcsname\relax\def\urlprefix{URL }\fi
\providecommand{\bibinfo}[2]{#2}
\providecommand{\eprint}[2][]{\url{#2}}

\bibitem[{\citenamefont{Goodman}(1996)}]{goodman}
\bibinfo{author}{\bibfnamefont{J.}~\bibnamefont{Goodman}},
  \emph{\bibinfo{title}{Introduction to Fourier Optics}}
  (\bibinfo{publisher}{McGraw Hill}, \bibinfo{address}{New York},
  \bibinfo{year}{1996}), \bibinfo{edition}{2nd} ed.

\bibitem[{\citenamefont{Born and Wolf}(1999)}]{born&wolf}
\bibinfo{author}{\bibfnamefont{M.}~\bibnamefont{Born}} \bibnamefont{and}
  \bibinfo{author}{\bibfnamefont{E.}~\bibnamefont{Wolf}},
  \emph{\bibinfo{title}{Principles of Optics}} (\bibinfo{publisher}{Cambridge
  University Press}, \bibinfo{address}{Cambridge}, \bibinfo{year}{1999}),
  \bibinfo{edition}{7th} ed.

\bibitem[{\citenamefont{Allen et~al.}(1992)\citenamefont{Allen, Beijersbergen,
  Spreeuw, and Woerdman}}]{allen1992pra}
\bibinfo{author}{\bibfnamefont{L.}~\bibnamefont{Allen}},
  \bibinfo{author}{\bibfnamefont{M.~W.} \bibnamefont{Beijersbergen}},
  \bibinfo{author}{\bibfnamefont{R.~J.~C.} \bibnamefont{Spreeuw}},
  \bibnamefont{and} \bibinfo{author}{\bibfnamefont{J.~P.}
  \bibnamefont{Woerdman}}, \bibinfo{journal}{Phys. Rev. A}
  \textbf{\bibinfo{volume}{45}}, \bibinfo{pages}{8185} (\bibinfo{year}{1992}).

\bibitem[{\citenamefont{Wang et~al.}(2012)\citenamefont{Wang, Yang, Fazal,
  Ahmed, Yan, Huang, Ren, Yue, Dolinar, Tur et~al.}}]{wang2012natphot}
\bibinfo{author}{\bibfnamefont{J.}~\bibnamefont{Wang}},
  \bibinfo{author}{\bibfnamefont{J.-Y.} \bibnamefont{Yang}},
  \bibinfo{author}{\bibfnamefont{I.~M.} \bibnamefont{Fazal}},
  \bibinfo{author}{\bibfnamefont{N.}~\bibnamefont{Ahmed}},
  \bibinfo{author}{\bibfnamefont{Y.}~\bibnamefont{Yan}},
  \bibinfo{author}{\bibfnamefont{H.}~\bibnamefont{Huang}},
  \bibinfo{author}{\bibfnamefont{Y.}~\bibnamefont{Ren}},
  \bibinfo{author}{\bibfnamefont{Y.}~\bibnamefont{Yue}},
  \bibinfo{author}{\bibfnamefont{S.}~\bibnamefont{Dolinar}},
  \bibinfo{author}{\bibfnamefont{M.}~\bibnamefont{Tur}}, \bibnamefont{et~al.},
  \bibinfo{journal}{Nat Photon} \textbf{\bibinfo{volume}{6}},
  \bibinfo{pages}{488} (\bibinfo{year}{2012}), ISSN \bibinfo{issn}{1749-4885}.

\bibitem[{\citenamefont{Bozinovic et~al.}(2013)\citenamefont{Bozinovic, Yue,
  Ren, Tur, Kristensen, Huang, Willner, and
  Ramachandran}}]{bozinovic2013science}
\bibinfo{author}{\bibfnamefont{N.}~\bibnamefont{Bozinovic}},
  \bibinfo{author}{\bibfnamefont{Y.}~\bibnamefont{Yue}},
  \bibinfo{author}{\bibfnamefont{Y.}~\bibnamefont{Ren}},
  \bibinfo{author}{\bibfnamefont{M.}~\bibnamefont{Tur}},
  \bibinfo{author}{\bibfnamefont{P.}~\bibnamefont{Kristensen}},
  \bibinfo{author}{\bibfnamefont{H.}~\bibnamefont{Huang}},
  \bibinfo{author}{\bibfnamefont{A.~E.} \bibnamefont{Willner}},
  \bibnamefont{and}
  \bibinfo{author}{\bibfnamefont{S.}~\bibnamefont{Ramachandran}},
  \bibinfo{journal}{Science} \textbf{\bibinfo{volume}{340}},
  \bibinfo{pages}{1545} (\bibinfo{year}{2013}), ISSN \bibinfo{issn}{0036-8075}.

\bibitem[{\citenamefont{Yan et~al.}(2014)\citenamefont{Yan, Xie, Lavery, Huang,
  Ahmed, Bao, Ren, Cao, Li, Zhao et~al.}}]{yan2014natcom}
\bibinfo{author}{\bibfnamefont{Y.}~\bibnamefont{Yan}},
  \bibinfo{author}{\bibfnamefont{G.}~\bibnamefont{Xie}},
  \bibinfo{author}{\bibfnamefont{M.~P.~J.} \bibnamefont{Lavery}},
  \bibinfo{author}{\bibfnamefont{H.}~\bibnamefont{Huang}},
  \bibinfo{author}{\bibfnamefont{N.}~\bibnamefont{Ahmed}},
  \bibinfo{author}{\bibfnamefont{C.}~\bibnamefont{Bao}},
  \bibinfo{author}{\bibfnamefont{Y.}~\bibnamefont{Ren}},
  \bibinfo{author}{\bibfnamefont{Y.}~\bibnamefont{Cao}},
  \bibinfo{author}{\bibfnamefont{L.}~\bibnamefont{Li}},
  \bibinfo{author}{\bibfnamefont{Z.}~\bibnamefont{Zhao}}, \bibnamefont{et~al.},
  \bibinfo{journal}{Nature Communications} \textbf{\bibinfo{volume}{5}},
  \bibinfo{pages}{4876} (\bibinfo{year}{2014}).

\bibitem[{\citenamefont{Karimipour et~al.}(2002)\citenamefont{Karimipour,
  Bahraminasab, and Bagherinezhad}}]{karimipour2002pra}
\bibinfo{author}{\bibfnamefont{V.}~\bibnamefont{Karimipour}},
  \bibinfo{author}{\bibfnamefont{A.}~\bibnamefont{Bahraminasab}},
  \bibnamefont{and}
  \bibinfo{author}{\bibfnamefont{S.}~\bibnamefont{Bagherinezhad}},
  \bibinfo{journal}{Phys. Rev. A} \textbf{\bibinfo{volume}{65}},
  \bibinfo{pages}{052331} (\bibinfo{year}{2002}).

\bibitem[{\citenamefont{Cerf et~al.}(2002)\citenamefont{Cerf, Bourennane,
  Karlsson, and Gisin}}]{cerf2002prl}
\bibinfo{author}{\bibfnamefont{N.~J.} \bibnamefont{Cerf}},
  \bibinfo{author}{\bibfnamefont{M.}~\bibnamefont{Bourennane}},
  \bibinfo{author}{\bibfnamefont{A.}~\bibnamefont{Karlsson}}, \bibnamefont{and}
  \bibinfo{author}{\bibfnamefont{N.}~\bibnamefont{Gisin}},
  \bibinfo{journal}{Phys. Rev. Lett.} \textbf{\bibinfo{volume}{88}},
  \bibinfo{pages}{127902} (\bibinfo{year}{2002}).

\bibitem[{\citenamefont{Nikolopoulos et~al.}(2006)\citenamefont{Nikolopoulos,
  Ranade, and Alber}}]{nikolopoulos2006pra}
\bibinfo{author}{\bibfnamefont{G.~M.} \bibnamefont{Nikolopoulos}},
  \bibinfo{author}{\bibfnamefont{K.~S.} \bibnamefont{Ranade}},
  \bibnamefont{and} \bibinfo{author}{\bibfnamefont{G.}~\bibnamefont{Alber}},
  \bibinfo{journal}{Phys. Rev. A} \textbf{\bibinfo{volume}{73}},
  \bibinfo{pages}{032325} (\bibinfo{year}{2006}).

\bibitem[{\citenamefont{Fujiwara et~al.}(2003)\citenamefont{Fujiwara, Takeoka,
  Mizuno, and Sasaki}}]{fujiwara2003prl}
\bibinfo{author}{\bibfnamefont{M.}~\bibnamefont{Fujiwara}},
  \bibinfo{author}{\bibfnamefont{M.}~\bibnamefont{Takeoka}},
  \bibinfo{author}{\bibfnamefont{J.}~\bibnamefont{Mizuno}}, \bibnamefont{and}
  \bibinfo{author}{\bibfnamefont{M.}~\bibnamefont{Sasaki}},
  \bibinfo{journal}{Phys. Rev. Lett.} \textbf{\bibinfo{volume}{90}},
  \bibinfo{pages}{167906} (\bibinfo{year}{2003}).

\bibitem[{\citenamefont{Cortese}(2004)}]{cortese2004pra}
\bibinfo{author}{\bibfnamefont{J.}~\bibnamefont{Cortese}},
  \bibinfo{journal}{Phys. Rev. A} \textbf{\bibinfo{volume}{69}},
  \bibinfo{pages}{022302} (\bibinfo{year}{2004}).

\bibitem[{\citenamefont{Ralph et~al.}(2007)\citenamefont{Ralph, Resch, and
  Gilchrist}}]{ralph2007pra}
\bibinfo{author}{\bibfnamefont{T.~C.} \bibnamefont{Ralph}},
  \bibinfo{author}{\bibfnamefont{K.~J.} \bibnamefont{Resch}}, \bibnamefont{and}
  \bibinfo{author}{\bibfnamefont{A.}~\bibnamefont{Gilchrist}},
  \bibinfo{journal}{Phys. Rev. A} \textbf{\bibinfo{volume}{75}},
  \bibinfo{pages}{022313} (\bibinfo{year}{2007}).

\bibitem[{\citenamefont{Lanyon et~al.}(2009)\citenamefont{Lanyon, Barbieri,
  Almeida, Jennewein, Ralph, Resch, Pryde, O/'Brien, Gilchrist, and
  White}}]{lanyon2009natphy}
\bibinfo{author}{\bibfnamefont{B.~P.} \bibnamefont{Lanyon}},
  \bibinfo{author}{\bibfnamefont{M.}~\bibnamefont{Barbieri}},
  \bibinfo{author}{\bibfnamefont{M.~P.} \bibnamefont{Almeida}},
  \bibinfo{author}{\bibfnamefont{T.}~\bibnamefont{Jennewein}},
  \bibinfo{author}{\bibfnamefont{T.~C.} \bibnamefont{Ralph}},
  \bibinfo{author}{\bibfnamefont{K.~J.} \bibnamefont{Resch}},
  \bibinfo{author}{\bibfnamefont{G.~J.} \bibnamefont{Pryde}},
  \bibinfo{author}{\bibfnamefont{J.~L.} \bibnamefont{O/'Brien}},
  \bibinfo{author}{\bibfnamefont{A.}~\bibnamefont{Gilchrist}},
  \bibnamefont{and} \bibinfo{author}{\bibfnamefont{A.~G.} \bibnamefont{White}},
  \bibinfo{journal}{Nat Phys} \textbf{\bibinfo{volume}{5}},
  \bibinfo{pages}{134} (\bibinfo{year}{2009}), ISSN \bibinfo{issn}{1745-2473}.

\bibitem[{\citenamefont{Jha et~al.}(2011)\citenamefont{Jha, Agarwal, and
  Boyd}}]{jha2011pra2}
\bibinfo{author}{\bibfnamefont{A.~K.} \bibnamefont{Jha}},
  \bibinfo{author}{\bibfnamefont{G.~S.} \bibnamefont{Agarwal}},
  \bibnamefont{and} \bibinfo{author}{\bibfnamefont{R.~W.} \bibnamefont{Boyd}},
  \bibinfo{journal}{Phys. Rev. A} \textbf{\bibinfo{volume}{83}},
  \bibinfo{pages}{053829} (\bibinfo{year}{2011}).

\bibitem[{\citenamefont{Kaszlikowski et~al.}(2000)\citenamefont{Kaszlikowski,
  Gnaci\ifmmode~\acute{n}\else \'{n}\fi{}ski, \ifmmode~\dot{Z}\else
  \.{Z}\fi{}ukowski, Miklaszewski, and Zeilinger}}]{kaszlikowski2000prl}
\bibinfo{author}{\bibfnamefont{D.}~\bibnamefont{Kaszlikowski}},
  \bibinfo{author}{\bibfnamefont{P.}~\bibnamefont{Gnaci\ifmmode~\acute{n}\else
  \'{n}\fi{}ski}},
  \bibinfo{author}{\bibfnamefont{M.}~\bibnamefont{\ifmmode~\dot{Z}\else
  \.{Z}\fi{}ukowski}},
  \bibinfo{author}{\bibfnamefont{W.}~\bibnamefont{Miklaszewski}},
  \bibnamefont{and}
  \bibinfo{author}{\bibfnamefont{A.}~\bibnamefont{Zeilinger}},
  \bibinfo{journal}{Phys. Rev. Lett.} \textbf{\bibinfo{volume}{85}},
  \bibinfo{pages}{4418} (\bibinfo{year}{2000}).

\bibitem[{\citenamefont{Collins et~al.}(2002)\citenamefont{Collins, Gisin,
  Linden, Massar, and Popescu}}]{collins2002prl}
\bibinfo{author}{\bibfnamefont{D.}~\bibnamefont{Collins}},
  \bibinfo{author}{\bibfnamefont{N.}~\bibnamefont{Gisin}},
  \bibinfo{author}{\bibfnamefont{N.}~\bibnamefont{Linden}},
  \bibinfo{author}{\bibfnamefont{S.}~\bibnamefont{Massar}}, \bibnamefont{and}
  \bibinfo{author}{\bibfnamefont{S.}~\bibnamefont{Popescu}},
  \bibinfo{journal}{Phys. Rev. Lett.} \textbf{\bibinfo{volume}{88}},
  \bibinfo{pages}{040404} (\bibinfo{year}{2002}).

\bibitem[{\citenamefont{V\'ertesi et~al.}(2010)\citenamefont{V\'ertesi,
  Pironio, and Brunner}}]{vetesi2010prl}
\bibinfo{author}{\bibfnamefont{T.}~\bibnamefont{V\'ertesi}},
  \bibinfo{author}{\bibfnamefont{S.}~\bibnamefont{Pironio}}, \bibnamefont{and}
  \bibinfo{author}{\bibfnamefont{N.}~\bibnamefont{Brunner}},
  \bibinfo{journal}{Phys. Rev. Lett.} \textbf{\bibinfo{volume}{104}},
  \bibinfo{pages}{060401} (\bibinfo{year}{2010}).

\bibitem[{\citenamefont{Leach et~al.}(2009)\citenamefont{Leach, Jack, Romero,
  Ritsch-Marte, Boyd, Jha, Barnett, Franke-Arnold, and
  Padgett}}]{leach2009optexp}
\bibinfo{author}{\bibfnamefont{J.}~\bibnamefont{Leach}},
  \bibinfo{author}{\bibfnamefont{B.}~\bibnamefont{Jack}},
  \bibinfo{author}{\bibfnamefont{J.}~\bibnamefont{Romero}},
  \bibinfo{author}{\bibfnamefont{M.}~\bibnamefont{Ritsch-Marte}},
  \bibinfo{author}{\bibfnamefont{R.~W.} \bibnamefont{Boyd}},
  \bibinfo{author}{\bibfnamefont{A.~K.} \bibnamefont{Jha}},
  \bibinfo{author}{\bibfnamefont{S.~M.} \bibnamefont{Barnett}},
  \bibinfo{author}{\bibfnamefont{S.}~\bibnamefont{Franke-Arnold}},
  \bibnamefont{and} \bibinfo{author}{\bibfnamefont{M.~J.}
  \bibnamefont{Padgett}}, \bibinfo{journal}{Opt. Express}
  \textbf{\bibinfo{volume}{17}}, \bibinfo{pages}{8287} (\bibinfo{year}{2009}).

\bibitem[{\citenamefont{Mair et~al.}(2001)\citenamefont{Mair, Vaziri, Weihs,
  and Zeilinger}}]{mair2001nature}
\bibinfo{author}{\bibfnamefont{A.}~\bibnamefont{Mair}},
  \bibinfo{author}{\bibfnamefont{A.}~\bibnamefont{Vaziri}},
  \bibinfo{author}{\bibfnamefont{G.}~\bibnamefont{Weihs}}, \bibnamefont{and}
  \bibinfo{author}{\bibfnamefont{A.}~\bibnamefont{Zeilinger}},
  \bibinfo{journal}{Nature} \textbf{\bibinfo{volume}{412}},
  \bibinfo{pages}{313} (\bibinfo{year}{2001}).

\bibitem[{\citenamefont{Gibson et~al.}(2004)\citenamefont{Gibson, Courtial,
  Padgett, Vasnetsov, Pas'ko, Barnett, and Franke-Arnold}}]{gibson2004optexp}
\bibinfo{author}{\bibfnamefont{G.}~\bibnamefont{Gibson}},
  \bibinfo{author}{\bibfnamefont{J.}~\bibnamefont{Courtial}},
  \bibinfo{author}{\bibfnamefont{M.~J.} \bibnamefont{Padgett}},
  \bibinfo{author}{\bibfnamefont{M.}~\bibnamefont{Vasnetsov}},
  \bibinfo{author}{\bibfnamefont{V.}~\bibnamefont{Pas'ko}},
  \bibinfo{author}{\bibfnamefont{S.~M.} \bibnamefont{Barnett}},
  \bibnamefont{and}
  \bibinfo{author}{\bibfnamefont{S.}~\bibnamefont{Franke-Arnold}},
  \bibinfo{journal}{Opt. Express} \textbf{\bibinfo{volume}{12}},
  \bibinfo{pages}{5448} (\bibinfo{year}{2004}).

\bibitem[{\citenamefont{Gruneisen et~al.}(2011)\citenamefont{Gruneisen, Dymale,
  Stoltenberg, and Steinhoff}}]{gruneisen2011njp}
\bibinfo{author}{\bibfnamefont{M.~T.} \bibnamefont{Gruneisen}},
  \bibinfo{author}{\bibfnamefont{R.~C.} \bibnamefont{Dymale}},
  \bibinfo{author}{\bibfnamefont{K.~E.} \bibnamefont{Stoltenberg}},
  \bibnamefont{and}
  \bibinfo{author}{\bibfnamefont{N.}~\bibnamefont{Steinhoff}},
  \bibinfo{journal}{New Journal of Physics} \textbf{\bibinfo{volume}{13}},
  \bibinfo{pages}{083030} (\bibinfo{year}{2011}).

\bibitem[{\citenamefont{Leach et~al.}(2002)\citenamefont{Leach, Padgett,
  Barnett, Franke-Arnold, and Courtial}}]{leach2002prl}
\bibinfo{author}{\bibfnamefont{J.}~\bibnamefont{Leach}},
  \bibinfo{author}{\bibfnamefont{M.~J.} \bibnamefont{Padgett}},
  \bibinfo{author}{\bibfnamefont{S.~M.} \bibnamefont{Barnett}},
  \bibinfo{author}{\bibfnamefont{S.}~\bibnamefont{Franke-Arnold}},
  \bibnamefont{and} \bibinfo{author}{\bibfnamefont{J.}~\bibnamefont{Courtial}},
  \bibinfo{journal}{Phys. Rev. Lett.} \textbf{\bibinfo{volume}{88}},
  \bibinfo{pages}{257901} (\bibinfo{year}{2002}).

\bibitem[{\citenamefont{Leach et~al.}(2004)\citenamefont{Leach, Courtial,
  Skeldon, Barnett, Franke-Arnold, and Padgett}}]{leach2004prl}
\bibinfo{author}{\bibfnamefont{J.}~\bibnamefont{Leach}},
  \bibinfo{author}{\bibfnamefont{J.}~\bibnamefont{Courtial}},
  \bibinfo{author}{\bibfnamefont{K.}~\bibnamefont{Skeldon}},
  \bibinfo{author}{\bibfnamefont{S.~M.} \bibnamefont{Barnett}},
  \bibinfo{author}{\bibfnamefont{S.}~\bibnamefont{Franke-Arnold}},
  \bibnamefont{and} \bibinfo{author}{\bibfnamefont{M.~J.}
  \bibnamefont{Padgett}}, \bibinfo{journal}{Phys. Rev. Lett.}
  \textbf{\bibinfo{volume}{92}}, \bibinfo{pages}{013601}
  (\bibinfo{year}{2004}).

\bibitem[{\citenamefont{Vasnetsov et~al.}(2003)\citenamefont{Vasnetsov, Torres,
  Petrov, and Torner}}]{vasnetsov2003optlett}
\bibinfo{author}{\bibfnamefont{M.}~\bibnamefont{Vasnetsov}},
  \bibinfo{author}{\bibfnamefont{J.}~\bibnamefont{Torres}},
  \bibinfo{author}{\bibfnamefont{D.}~\bibnamefont{Petrov}}, \bibnamefont{and}
  \bibinfo{author}{\bibfnamefont{L.}~\bibnamefont{Torner}},
  \bibinfo{journal}{Optics letters} \textbf{\bibinfo{volume}{28}},
  \bibinfo{pages}{2285} (\bibinfo{year}{2003}).

\bibitem[{\citenamefont{Courtial et~al.}(1998)\citenamefont{Courtial,
  Robertson, Dholakia, Allen, and Padgett}}]{courtial1998prl}
\bibinfo{author}{\bibfnamefont{J.}~\bibnamefont{Courtial}},
  \bibinfo{author}{\bibfnamefont{D.~A.} \bibnamefont{Robertson}},
  \bibinfo{author}{\bibfnamefont{K.}~\bibnamefont{Dholakia}},
  \bibinfo{author}{\bibfnamefont{L.}~\bibnamefont{Allen}}, \bibnamefont{and}
  \bibinfo{author}{\bibfnamefont{M.~J.} \bibnamefont{Padgett}},
  \bibinfo{journal}{Phys. Rev. Lett.} \textbf{\bibinfo{volume}{81}},
  \bibinfo{pages}{4828} (\bibinfo{year}{1998}).

\bibitem[{\citenamefont{Berkhout et~al.}(2010)\citenamefont{Berkhout, Lavery,
  Courtial, Beijersbergen, and Padgett}}]{berkhout2010prl}
\bibinfo{author}{\bibfnamefont{G.~C.} \bibnamefont{Berkhout}},
  \bibinfo{author}{\bibfnamefont{M.~P.} \bibnamefont{Lavery}},
  \bibinfo{author}{\bibfnamefont{J.}~\bibnamefont{Courtial}},
  \bibinfo{author}{\bibfnamefont{M.~W.} \bibnamefont{Beijersbergen}},
  \bibnamefont{and} \bibinfo{author}{\bibfnamefont{M.~J.}
  \bibnamefont{Padgett}}, \bibinfo{journal}{Phys. Rev. Lett.}
  \textbf{\bibinfo{volume}{105}}, \bibinfo{pages}{153601}
  (\bibinfo{year}{2010}).

\bibitem[{\citenamefont{Lavery et~al.}(2012)\citenamefont{Lavery, Robertson,
  Berkhout, Love, Padgett, and Courtial}}]{lavery2012optexp}
\bibinfo{author}{\bibfnamefont{M.~P.} \bibnamefont{Lavery}},
  \bibinfo{author}{\bibfnamefont{D.~J.} \bibnamefont{Robertson}},
  \bibinfo{author}{\bibfnamefont{G.~C.} \bibnamefont{Berkhout}},
  \bibinfo{author}{\bibfnamefont{G.~D.} \bibnamefont{Love}},
  \bibinfo{author}{\bibfnamefont{M.~J.} \bibnamefont{Padgett}},
  \bibnamefont{and} \bibinfo{author}{\bibfnamefont{J.}~\bibnamefont{Courtial}},
  \bibinfo{journal}{Optics express} \textbf{\bibinfo{volume}{20}},
  \bibinfo{pages}{2110} (\bibinfo{year}{2012}).

\bibitem[{\citenamefont{Lightman et~al.}(2017)\citenamefont{Lightman, Hurvitz,
  Gvishi, and Arie}}]{lightman2017optica}
\bibinfo{author}{\bibfnamefont{S.}~\bibnamefont{Lightman}},
  \bibinfo{author}{\bibfnamefont{G.}~\bibnamefont{Hurvitz}},
  \bibinfo{author}{\bibfnamefont{R.}~\bibnamefont{Gvishi}}, \bibnamefont{and}
  \bibinfo{author}{\bibfnamefont{A.}~\bibnamefont{Arie}},
  \bibinfo{journal}{Optica} \textbf{\bibinfo{volume}{4}}, \bibinfo{pages}{605}
  (\bibinfo{year}{2017}).

\bibitem[{\citenamefont{Bryngdahl}(1974{\natexlab{a}})}]{bryngdahl1974josa}
\bibinfo{author}{\bibfnamefont{O.}~\bibnamefont{Bryngdahl}},
  \bibinfo{journal}{JOSA} \textbf{\bibinfo{volume}{64}}, \bibinfo{pages}{1092}
  (\bibinfo{year}{1974}{\natexlab{a}}).

\bibitem[{\citenamefont{Bryngdahl}(1974{\natexlab{b}})}]{bryngdahl1974optcomm}
\bibinfo{author}{\bibfnamefont{O.}~\bibnamefont{Bryngdahl}},
  \bibinfo{journal}{Optics Communications} \textbf{\bibinfo{volume}{10}},
  \bibinfo{pages}{164} (\bibinfo{year}{1974}{\natexlab{b}}).

\bibitem[{\citenamefont{Yao et~al.}(2006)\citenamefont{Yao, Franke-Arnold,
  Courtial, Barnett, and Padgett}}]{yao2006opex}
\bibinfo{author}{\bibfnamefont{E.}~\bibnamefont{Yao}},
  \bibinfo{author}{\bibfnamefont{S.}~\bibnamefont{Franke-Arnold}},
  \bibinfo{author}{\bibfnamefont{J.}~\bibnamefont{Courtial}},
  \bibinfo{author}{\bibfnamefont{S.}~\bibnamefont{Barnett}}, \bibnamefont{and}
  \bibinfo{author}{\bibfnamefont{M.}~\bibnamefont{Padgett}},
  \bibinfo{journal}{Opt. Express} \textbf{\bibinfo{volume}{14}},
  \bibinfo{pages}{9071} (\bibinfo{year}{2006}).

\bibitem[{\citenamefont{Jha et~al.}(2008)\citenamefont{Jha, Jack, Yao, Leach,
  Boyd, Buller, Barnett, Franke-Arnold, and Padgett}}]{jha2008pra2}
\bibinfo{author}{\bibfnamefont{A.~K.} \bibnamefont{Jha}},
  \bibinfo{author}{\bibfnamefont{B.}~\bibnamefont{Jack}},
  \bibinfo{author}{\bibfnamefont{E.}~\bibnamefont{Yao}},
  \bibinfo{author}{\bibfnamefont{J.}~\bibnamefont{Leach}},
  \bibinfo{author}{\bibfnamefont{R.~W.} \bibnamefont{Boyd}},
  \bibinfo{author}{\bibfnamefont{G.~S.} \bibnamefont{Buller}},
  \bibinfo{author}{\bibfnamefont{S.~M.} \bibnamefont{Barnett}},
  \bibinfo{author}{\bibfnamefont{S.}~\bibnamefont{Franke-Arnold}},
  \bibnamefont{and} \bibinfo{author}{\bibfnamefont{M.~J.}
  \bibnamefont{Padgett}}, \bibinfo{journal}{Phys. Rev. A}
  \textbf{\bibinfo{volume}{78}}, \bibinfo{eid}{043810}
  (pages~\bibinfo{numpages}{4}) (\bibinfo{year}{2008}).

\bibitem[{\citenamefont{Jha et~al.}(2010)\citenamefont{Jha, Leach, Jack,
  Franke-Arnold, Barnett, Boyd, and Padgett}}]{jha2010prl}
\bibinfo{author}{\bibfnamefont{A.~K.} \bibnamefont{Jha}},
  \bibinfo{author}{\bibfnamefont{J.}~\bibnamefont{Leach}},
  \bibinfo{author}{\bibfnamefont{B.}~\bibnamefont{Jack}},
  \bibinfo{author}{\bibfnamefont{S.}~\bibnamefont{Franke-Arnold}},
  \bibinfo{author}{\bibfnamefont{S.~M.} \bibnamefont{Barnett}},
  \bibinfo{author}{\bibfnamefont{R.~W.} \bibnamefont{Boyd}}, \bibnamefont{and}
  \bibinfo{author}{\bibfnamefont{M.~J.} \bibnamefont{Padgett}},
  \bibinfo{journal}{Phys. Rev. Lett.} \textbf{\bibinfo{volume}{104}},
  \bibinfo{pages}{010501} (\bibinfo{year}{2010}).

\bibitem[{\citenamefont{Arlt et~al.}(2001)\citenamefont{Arlt, Kuhn, and
  Dholakia}}]{arlt2001jmo}
\bibinfo{author}{\bibfnamefont{J.}~\bibnamefont{Arlt}},
  \bibinfo{author}{\bibfnamefont{R.}~\bibnamefont{Kuhn}}, \bibnamefont{and}
  \bibinfo{author}{\bibfnamefont{K.}~\bibnamefont{Dholakia}},
  \bibinfo{journal}{journal of modern optics} \textbf{\bibinfo{volume}{48}},
  \bibinfo{pages}{783} (\bibinfo{year}{2001}).

\bibitem[{\citenamefont{Arriz\'{o}n et~al.}(2007)\citenamefont{Arriz\'{o}n,
  Ruiz, Carrada, and Gonz\'{a}lez}}]{arrizon2007josaa}
\bibinfo{author}{\bibfnamefont{V.}~\bibnamefont{Arriz\'{o}n}},
  \bibinfo{author}{\bibfnamefont{U.}~\bibnamefont{Ruiz}},
  \bibinfo{author}{\bibfnamefont{R.}~\bibnamefont{Carrada}}, \bibnamefont{and}
  \bibinfo{author}{\bibfnamefont{L.~A.} \bibnamefont{Gonz\'{a}lez}},
  \bibinfo{journal}{J. Opt. Soc. Am. A} \textbf{\bibinfo{volume}{24}},
  \bibinfo{pages}{3500} (\bibinfo{year}{2007}).

\bibitem[{\citenamefont{Mirhosseini et~al.}(2013)\citenamefont{Mirhosseini,
  Malik, Shi, and Boyd}}]{mirhosseini2013natcomm}
\bibinfo{author}{\bibfnamefont{M.}~\bibnamefont{Mirhosseini}},
  \bibinfo{author}{\bibfnamefont{M.}~\bibnamefont{Malik}},
  \bibinfo{author}{\bibfnamefont{Z.}~\bibnamefont{Shi}}, \bibnamefont{and}
  \bibinfo{author}{\bibfnamefont{R.~W.} \bibnamefont{Boyd}},
  \bibinfo{journal}{Nature communications} \textbf{\bibinfo{volume}{4}}
  (\bibinfo{year}{2013}).

\bibitem[{\citenamefont{Huang et~al.}(2015)\citenamefont{Huang, Milione,
  Lavery, Xie, Ren, Cao, Ahmed, Nguyen, Nolan, Li et~al.}}]{huang2015scirep}
\bibinfo{author}{\bibfnamefont{H.}~\bibnamefont{Huang}},
  \bibinfo{author}{\bibfnamefont{G.}~\bibnamefont{Milione}},
  \bibinfo{author}{\bibfnamefont{M.~P.} \bibnamefont{Lavery}},
  \bibinfo{author}{\bibfnamefont{G.}~\bibnamefont{Xie}},
  \bibinfo{author}{\bibfnamefont{Y.}~\bibnamefont{Ren}},
  \bibinfo{author}{\bibfnamefont{Y.}~\bibnamefont{Cao}},
  \bibinfo{author}{\bibfnamefont{N.}~\bibnamefont{Ahmed}},
  \bibinfo{author}{\bibfnamefont{T.~A.} \bibnamefont{Nguyen}},
  \bibinfo{author}{\bibfnamefont{D.~A.} \bibnamefont{Nolan}},
  \bibinfo{author}{\bibfnamefont{M.-J.} \bibnamefont{Li}},
  \bibnamefont{et~al.}, \bibinfo{journal}{Scientific reports}
  \textbf{\bibinfo{volume}{5}}, \bibinfo{pages}{14931} (\bibinfo{year}{2015}).

\bibitem[{\citenamefont{O'Sullivan et~al.}(2012)\citenamefont{O'Sullivan,
  Mirhosseini, Malik, and Boyd}}]{osullivan2012optexp}
\bibinfo{author}{\bibfnamefont{M.~N.} \bibnamefont{O'Sullivan}},
  \bibinfo{author}{\bibfnamefont{M.}~\bibnamefont{Mirhosseini}},
  \bibinfo{author}{\bibfnamefont{M.}~\bibnamefont{Malik}}, \bibnamefont{and}
  \bibinfo{author}{\bibfnamefont{R.~W.} \bibnamefont{Boyd}},
  \bibinfo{journal}{Optics express} \textbf{\bibinfo{volume}{20}},
  \bibinfo{pages}{24444} (\bibinfo{year}{2012}).

\bibitem[{\citenamefont{Romero and Dickey}(2007)}]{romero2007josaa}
\bibinfo{author}{\bibfnamefont{L.~A.} \bibnamefont{Romero}} \bibnamefont{and}
  \bibinfo{author}{\bibfnamefont{F.~M.} \bibnamefont{Dickey}},
  \bibinfo{journal}{JOSA A} \textbf{\bibinfo{volume}{24}},
  \bibinfo{pages}{2280} (\bibinfo{year}{2007}).

\end{thebibliography}

%
%\section*{Disclosures}
%For \textit{Biomedical Optics Express} submissions only, disclosures should be listed in a separate nonnumbered section at the end of the manuscript. List the Disclosures codes identified on OSA's \href{http://www.osapublishing.org/submit/review/conflicts-interest-policy.cfm}{Conflict of Interest policy page}, as shown in the examples below:\\
%\\
%ABC: 123 Corporation (I,E,P), DEF: 456 Corporation (R,S). GHI: 789 Corporation (C).\\
%\\
%If there are no disclosures, then list ``The authors declare that there are no conflicts of interest related to this article.''

\end{document}